# De-carbonization of global energy use during the COVID-19 pandemic

Zhu Liu[1*†], Biqing Zhu[1†], Philippe Ciais[2], Steven J. Davis[3], Chenxi Lu[1], Haiwang Zhong[1], Piyu Ke[1], Yanan Cui[1], Zhu Deng[1], Duo Cui[1], Taochun Sun[1], Xinyu Dou[1], Jianguang Tan[1], Rui Guo[1], Bo Zheng[2], Katsumasa Tanaka[2], Wenli Zhao[4], Pierre Gentine[4]

1 Department of Earth System Science, Tsinghua University, Beijing, China.
2 Laboratoire des Sciences du Climate et de l'Environnement LSCE, Orme de Merisiers, Gif-sur-Yvette, France
3 Department of Earth System Science, University of California, Irvine, 3232 Croul Hall, Irvine, CA, USA
4 Department of Earth & Environment Sciences, Columbia University

Corresponding author: zhuliu@tsinghua.edu.cn

# Abstract

**The COVID-19 pandemic has disrupted human activities, leading to unprecedented decreases in both global energy demand and GHG emissions. Yet a little known that there is also a "low carbon" shift of the global energy system in 2020. Here, using the near-real-time data on energy-related GHG emissions from 30 countries (~70% of global power generation), we show that the pandemic caused an unprecedented de-carbonization of global power system, representing by a dramatic decrease in the carbon intensity of power sector that reached a historical low of 414.9 tCO2eq/GWh in 2020. Moreover, the share of energy derived from renewable and low-carbon sources (nuclear, hydro-energy, wind, solar, geothermal, and biomass) exceeded that from coal and oil for the first time in history in May of 2020. The decrease in global net energy demand (-1.3% in the first half of 2020 relative to the average of the period in 2016-2019) masks a large down-regulation of fossil-fuel-burning power plants supply (-6.1%) coincident with a surge of low-carbon sources (+6.2%). Concomitant changes in the diurnal cycle of electricity demand also favored low-carbon generators, including a flattening of the morning ramp, a lower midday peak, and delays in both the morning and midday load peaks in most countries. However, emission intensities in the power sector have since rebounded in many countries, and a key question for climate mitigation is thus to what extent countries can achieve and maintain lower, pandemic-level carbon intensities of electricity as part of a green recovery.**

***[245 words]***

Since late 2019, the ongoing coronavirus disease (COVID-19) has caused more than a million deaths worldwide[1] . This has led to the largest disruption of human activities since World War II, with drastic changes in energy consumption and generation patterns, leading to an unprecedented drop of $CO_2$ emissions[2-5]. Moreover, the relative decrease of $CO_2$ emissions from the electricity sector was disproportionately greater than the decreases in electricity demand and generation, pointing to a reduction in carbon intensity of the energy sector. For example, electricity use in China between January and April of 2020 was 4.7%[5] lower than during the same months of 2019, but the corresponding decrease in electricity-related emissions was 6.0%[6]. In particular, coal power was more negatively affected by the pandemic than other sources. The observed vulnerability of coal power to the negative demand shock of COVID-19 in many countries leads us to speculate that this sector may be close to a tipping point, beyond which substantial coal-fired generation capacity may become economically disadvantaged. This could hasten the phasing-out of coal power plants worldwide[7]. Changes in electricity demand and supply during the pandemic may thus have important lessons for near-term reductions in energy-related $CO_2$ emissions, but the necessary in-depth analysis is yet lacking.

High spatial- and temporal- source specific electricity data can help understand why de-carbonization occurred and can help interpret the short-term immediate and long term effect COVID-19 has on the electricity sector. Traditionally, this sector tend to be analyzed either in aggregated time scale (monthly or yearly, often updated with time-lag) or for refined spatial locations (national or regional) [3]. While monthly or yearly aggregated data do provide knowledge base for evaluating the overall profiles of the electricity sector, information on day-to-day and diurnal electricity consumption and generation patterns induced by human behavior is unfortunately omitted. Regional daily electricity profiles can fill in some of this gap by providing snapshots of certain specific locations. However, it fails in providing a dynamic integral portrait of the global electricity system. A global scaled dataset on source specific electricity generation and consumption, with high time frequency and with high spatial coverage is therefore urgently needed to understand what exactly changes electricity sector has gone through during the pandemic globally, and when and how these changes happened.

Here, we compiled and analyzed near-real-time, sub-hourly, sub-regionally electricity demand, electricity generation and related emissions from January of 2016 to June of 2020[8] (see also: http://carbonmonitor.org). Specifically, we quantify the patterns and drivers of global carbon intensity change in power generation (defined as the amount of $CO_2$-equivalent greenhouse gases emitted per unit energy generated) since the onset of the COVID-19 pandemic. We accomplished this by providing a long-term dataset revealing the historical development and daily variations of the carbon intensity and the energy structure change (defined as the proportion of electricity generated by each energy source) in global power generation, (2) examining the drivers of the energy structure change during the pandemic by assessing the driving force: the electricity demand pattern shift in the corresponding time period, (3) evaluating how the newly formed electricity demand pattern since the COVID-19 pandemic facilitated the energy structure change by unfolding the diurnal pattern shift of each energy source, (4) assessing the reversed process by evaluating the carbon intensity and $CO_2$ emission rebound from the electricity sector with the temporary recovery from lockdown in summer 2020. We included all major electricity generation sources in this study, covering electricity supplied by thermal power of oil, natural gas and coal, as well as, nuclear, hydro, wind, solar and other renewables (mainly bioenergy and geothermal energy). We compiled sub-hourly, hourly and daily data from 30 countries that account for ~68% of global electricity generation and global carbon emissions (See Supplementary Table S1).

**Accelerated decline of global carbon intensity marked by declined fossil electricity and more renewables**

During the first half of 2020, the carbon intensity of electricity generated worldwide continued its fast decline and reached a historical low of 414.9 tCO2eq/GWh (Fig1a). In May of 2020, the share of low-carbon sources in the global power mix reached an historical high of 47%, reflecting increased shares of electricity from low-carbon sources and decreased shares from fossil sources (Figs. 1b and 1c). Some of the increases in wind and solar generation were related to exceptional weather conditions in some regions; nuclear and hydro remained at the level of previous years, and biomass together with geothermal decreased (Fig. 1b). Altogether, the share of low-carbon sources worldwide surpassed power generated from coal and oil combined for the first time (Fig. 1b), growing by 2.4% in 2020--almost three times faster than the average 2014-2019 (Fig. 1d). As a result, these changes led to a record decline of GHG emission in the power sector by 24.2% (Fig. 1d; hereafter all reported changes refer to the period of January 1st till July 1st relative to the same period in previous years).

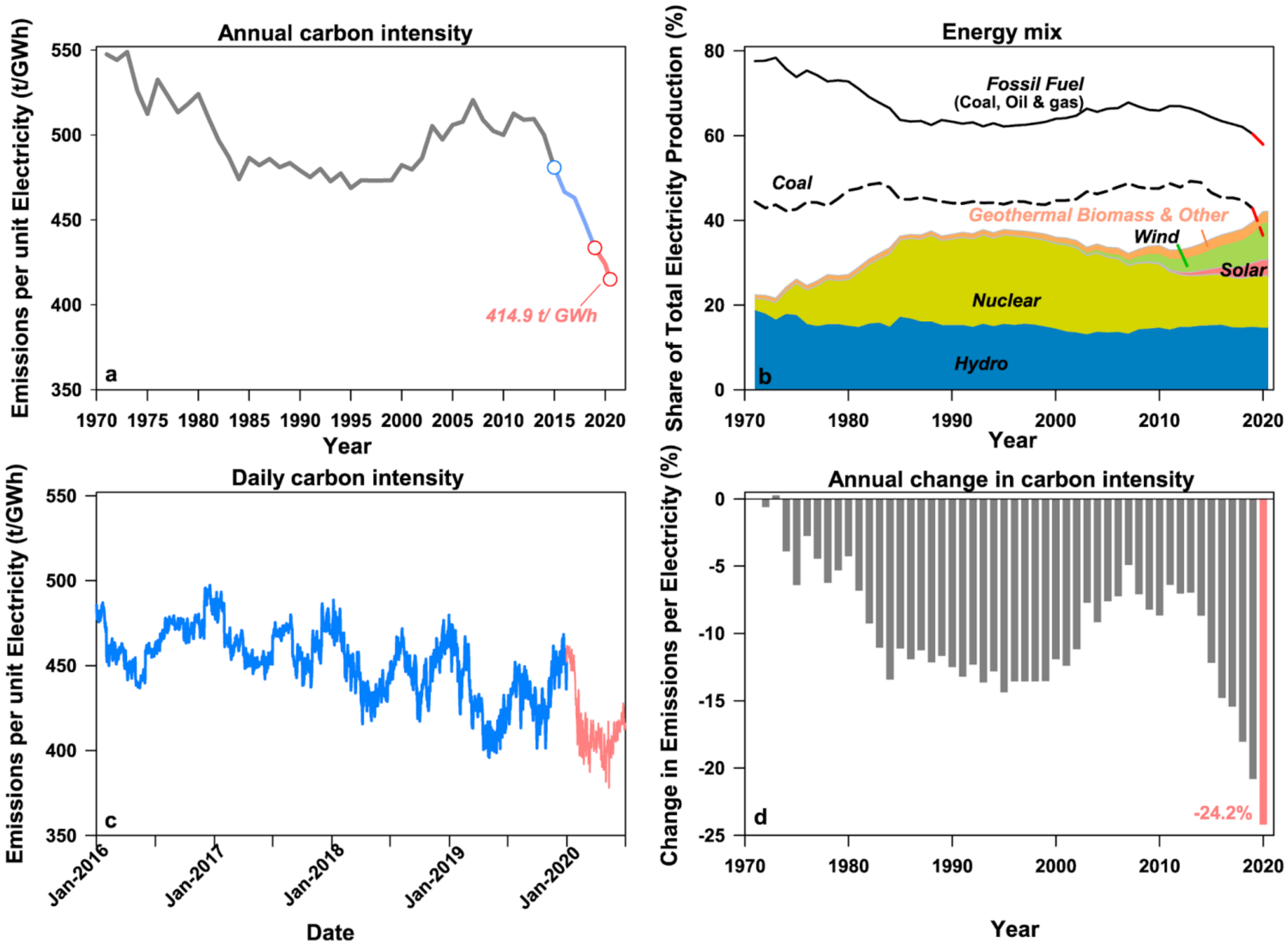

**Figure 1 | Carbon intensity and share of energy source change in the global power generation system.** Fig1.a shows the carbon intensity (defined as carbon emissions per unit energy) for annual carbon intensity from 1971 to 2020. Black lines represent annual carbon intensity 1971 and 2015; Blue lines represent annual carbon intensity between 2016 and 2019; red lines represent annual carbon intensity in 2020 (from 1st January to 1st July). Fig1.b shows the contribution of different energy sources to total energy production. Low carbon sources are shown in shaded areas, thermal sources are shown with lines. Red lines represent the change since 1st January, 2020. Fig1. c shows the daily carbon intensity from 1st January, 2016 to 1st July, 2020. Blue lines represent daily carbon intensity from 2016 to 2019; red lines represent daily carbon intensity in 2020. The relative change in carbon intensity is shown in Fig1.d, with 1971 as reference. The change in carbon intensity in year i is calculated as the difference of carbon intensity in year i and in 1971. Red bar represents the numbers in year 2020.

Electricity demand is sensitive to seasonal temperature variations for winter heating in cold countries and summer air conditioning in many subtropical and tropical countries[9]. Yet, the share of electricity in the final energy use for heating varies substantially across countries, e.g., from 1% in Poland to 18% in Spain. In addition to spurring electricity demand for air conditioning, hot conditions also reduce the efficiency of thermal power plant operations, so that high temperatures also directly affect the mix of power sources. The first few winter months of 2020 were exceptionally warm in many places[10] (Supplementary Table S5), leading to a reduction of heating demand on top of the COVID-19 impacts. During 2020 winter time (from January to March), out of the total energy generation decrease of 5.39% (compared to 2019), we attributed 0.72% to abnormally warm temperatures (Least-Squares Regression Model, see *Methods*). Out of all countries analyzed, we found medium to strong correlation between daily electricity generation and daily average temperature in 18 countries, while taking into account public holidays (modeled separately) and population density (temperature was first weighted by population before modeling; country-specific results see Supplementary Table S4)

In order to filter seasonal temperature variations while retaining inter-annual temperature anomalies, we first compiled the baseline electricity generation using data from 2016-2019 for each component of the global power mix in Figure 2 (dashed lines and shaded areas, see also Supplementary Fig. S1). Both fossil and nuclear which deliver the base power supply show a double seasonality with a peak in January and in July: the summer peak being smaller than the winter one for oil (Fig 2d) and nuclear (Fig 2e), but higher coal (Fig 2b) and gas (Fig 2c), which is mainly related to the exposure of countries with a different energy mix to heating and cooling demands inferred from heating and cooling degree days (HDD[11] and CDD[12]). Solar energy follows the solar radiation pattern in the northern hemisphere (Fig 2h), and hydro energy peaks in summer (Fig 2f), mainly owing to a higher production in monsoon Asian countries[13]. Wind energy exhibits a broad maximum from the autumn to the spring equinox, consistent with the prevailing intensity of the zonal atmospheric circulation in the northern hemisphere (Fig 2g). In the past five years, wind and solar showed a steady increase, while other sources remained stable (Supplementary Fig. S1).

Then we compared the daily source specific electricity generation in 2020 (Fig 2, solid lines) to the baseline values to compute the corresponding de-seasoned change. In the first six months of year 2020, global power generation decreased by -1.5% (-121.2 TWh, Fig2a), Noticeably, Japan, the US and Europe experienced the greatest relative declines of -11.8% (56.3 TWh), -8.5% (166.9 TWh) and -5.6% (80.1 TWh), respectively. In contrast, there was a net increase in China (+7.6%, 231.5TWh) because the COVID-induced declines in February coincided with the timing of the annual Spring Festival when the power supply is normally reduced (-0.3%, -1.5 TWh)

The development of different energy sources, however, were impacted differently by the pandemic in 2020. In the first half of 2020, the absolute amount of wind and solar power showed large positive anomalies, +30.5% and +49.6%, explained by very windy conditions (in February in Western Europe) and very sunny springs (in both Europe and US). In contrast, Figure 2b shows that coal power declined by the end of January at the onset of the pandemic China and returned close to the monthly average by June (Supplementary Fig. S2). Oil power had remained low since Jan 2020 despite the low prices. The share of electricity from gas was higher than average before March, favored by low market prices of gas in the U.S. and Europe allowing to switch rapidly to a higher gas share. Later, global gas power decreased to reach within its monthly inter-annual variability from April to June (Fig 2c and Supplementary Table S4).

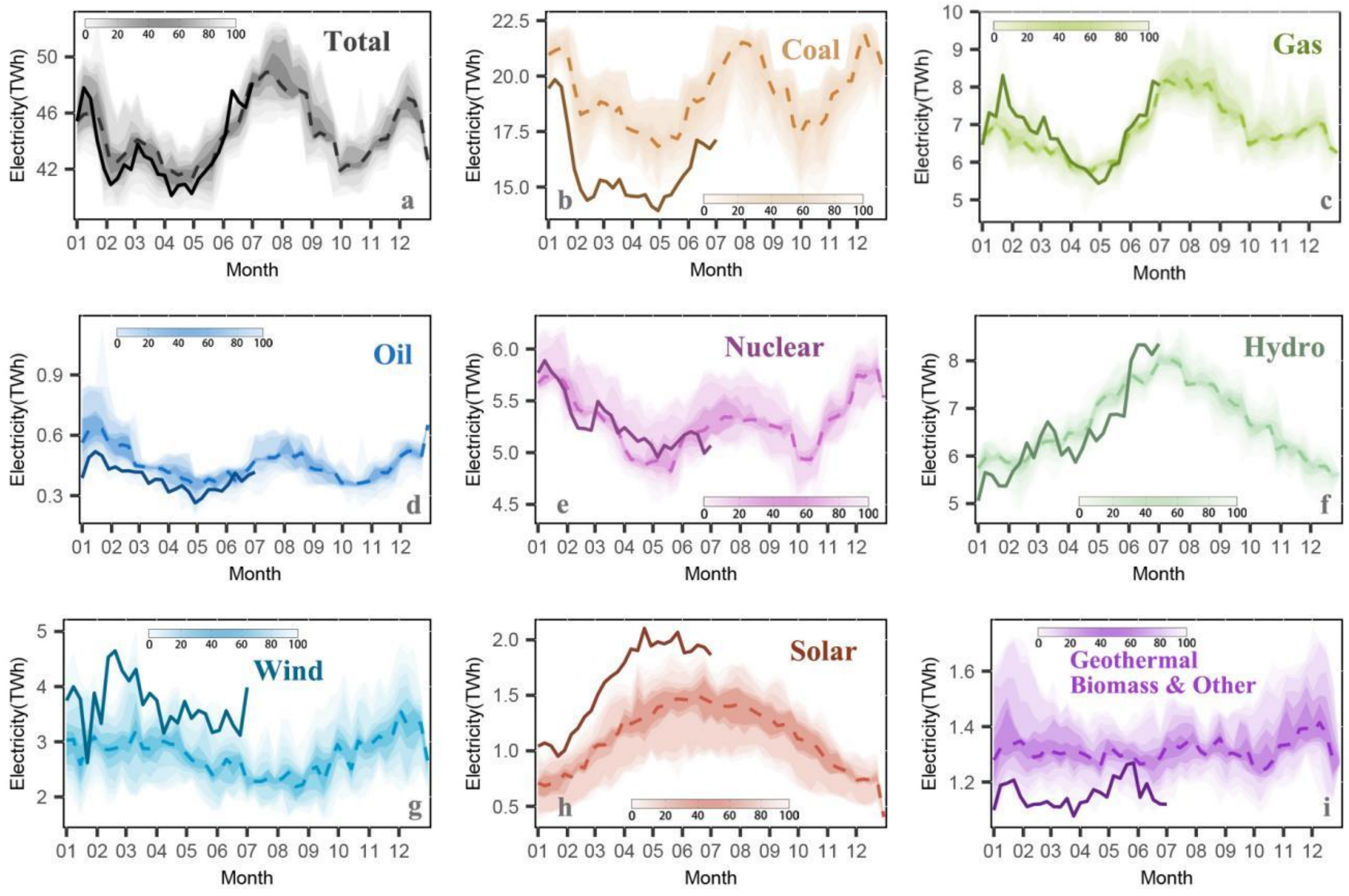

**Figure 2 | Daily global total and source-wise power generation.** Solid line indicates the daily trend of (a)total power generation, (b) coal power generation, (c) gas power generation, (d) oil power generation, (e) nuclear power generation, (f) hydro power generation, (g) wind power generation (h) solar power generation, and (i) geothermal, biomass&other power generation, since 1st January, 2020. Shaded areas show the range of daily power generation for the reference period (from 1st January, 2016 to 31st December, 2019). Darkness of shading color corresponds to the confidence interval from uncertainty analyses. Dashed lines refer to the medium of reference data. This figure presents the trends for global power generation, including data from Europe (EU27&UK), US, China, India, Brazil and Japan. For trends in other countries and regions, see Supplementary Fig. S1 and Supplementary Fig. S2.

- Declines in fossil electricity generation

In the first six months of 2020, electricity from fossil fuel sources decreased by -6.1% (292.2 TWh Supplementary Fig S2), largely reflecting an -8.7% (-310.58 TWh) decrease in coal electricity (Fig2b). The decline was caused by a combination of factors, including a negative trend of coal in the energy mixes of Europe (-39.1%), the U.S. (-41.5%), and Japan (-9.8%), against the background of sharp declines in economic activity due to public health responses to the COVID-19 pandemic, particularly in industrial production (e.g., industrial demand declined by -5.2%, or -108.3 TWh in China, by -14.3%, or -73.7 TWh, in India, and by -20.8%, or -2.1 TWh, in Brazil; Supplementary Fig. S2).

Electricity from oil also began 2020 at lower than past year levels (-40.1%, -7.5 TWh in January), and these lows continued during the pandemic, reflecting policies aimed at reducing the use of this fuel in many countries[14,15], with a total decline of 35.6% in the first 6 months of 2020 (76.5 TWh $\pm$ 18.9%). In January, lower shares of electricity from oil were observed in Europe (-36.3%), the U.S. (-12.6%), Japan (-13.0%), India (-48.0%) and Brazil (-65.8%), but oil electricity increased in China (+3.7% ). However, during February and March, when China was under strict lockdown, Chinese electricity from oil did decline by about -8% relative to the reference period.

In contrast to coal and oil, 2020 began with an increase in electricity supplied by natural gas (+4.8%, +10.3 TWh growth in January, +9.2%, +12.2 TWh in February Fig2c), driven by rising gas use in Europe (+13.3%, +6.46 TWh in January), the U.S. (+25.7%, +23.6 TWh in February), and Brazil (+52.2%, +2.0 TWh in January). However, the growth in gas use was cut short by the pandemic. Electricity from gas decreased with lockdowns in the U.S., Europe, and Brazil (e.g., the reduced operation time of natural gas energy operation systems) (Fig. 2; Supplementary Fig. S2).

- More renewables in the first half of 2020

We observe a substantial increase in global wind (+29.4% increase from $5.17\times10^2$ TWh $\pm$ 15% average of 2016-2019， Fig2g) and an even larger increase in solar power generation (+48.0% increase from $1.98\times10^2$ TWh $\pm$ 24.7%, average of 2016-2019, Fig2h). Meanwhile, there were small decreases in generation from hydropower, biomass and other renewables, while global nuclear was similar as reference years. Altogether, the share of low- carbon sources in the energy mix reached an historical high of 47% in May of 2020.

The strongest growth among all energy sources was seen in solar power, with a 49.4% increase in January 2020, and a +52.5% increase in April. Furthermore, the increase of solar energy was observed in every region/country analyzed and every month from January to June. The largest increase was in China, with a rise of +60.8% (6.2 TWh) in January and of +88.4% (10.9 TWh) in April. Brazil and Japan also show large solar energy growth (+221.3%, 1.7 TWh and +31.2%, 8.83 TWh respectively). Increases are smaller but still substantial in Europe and the U.S. (+25.6%, 14.1 TWh and +16.3%, 5.3 TWh, respectively).

Wind power was already 21.0 % higher than the reference (91.5 TWh $\pm$ 19.5%, 2016-2019 average, Fig2g) in January 2020, mainly due to continuously increasing construction of wind power generation and power purchasing activities in Europe, US and China[15] (Supplementary Fig. S2). After a brief drop back to the level as previous years at the end of January, global electricity from wind grew strongly throughout the first half of year 2020, by a total of 29.0% (1.5 $\times 10^2$ TWh). Wind energy increased in every region/country analyzed in this study, though to different extents. The largest increase is in China (+41.9%, 67.9 TWh), followed by Europe (+29.4%, 51.9 TWh) and the U.S. (+23.1%, 31.9 TWh).

In contrast to wind and solar energy sources, hydro-power began 2020 lower than the reference period (-8.4% in January, compared to 1.82 ×102 TWh $\pm$ 2.7%, Fig2f), but reached around the reference level from February untill June. Geothermal, biomass and other renewables have shown a 14.6% decline since the beginning of year 2020 with no obvious change during the COVID-19 pandemic.

- Contrasting regional development of Nuclear energy

Global nuclear energy did not show large changes during the first half of 2020 (9.58 ×102 TWh, compared to 9.65 ×102 TWh $\pm$ 4.4%, Fig2e). However, there is a remarkable difference between countries. In Europe (-12.1%), the U.S. (-3.2%), and Brazil (-7.3%), the shares of nuclear electricity declined, while in China (+36.8%), India (+9.0%), and Japan (+60.8%), nuclear power generation increased in the first half of 2020 (Supplementary Fig. S1). Nuclear power generation decline sharply in China in February following the lockdown measures. However, no clear evidence of impact from COVID-19 was observed in the nuclear energy production in other countries (Supplementary Fig S3).

**Declined and shifted energy demand pattern following the COVID-19 pandemic**

The global electricity demand and generation systems were immediately impacted by the COVID-19 pandemic, and this impact induced change is not homogeneous throughout the day. By comparing the diurnal cycle of power generation before and after COVID-19, with the same time period as reference, we found 1) a flattening of morning ramp (the average power delivered per hour before the peak in midday) 2) a lower peak demand, and 3) a delay in the peak time from around 8 AM before lockdown (week 11) to as late as around 12 PM after lock downs (Fig. 3). For Europe and the U.S., both morning demand peaks declined and peaks were delayed (Fig. 3). In the US, the evening peak surpassed or equaled the morning one, whereas in Europe, the evening peak load declined to a similar level as the morning peak.

The change of diurnal power profiles changed to a more "weekend-like" pattern after the lockdown. Prior to lockdown, the 2020 weekday energy demand was very close to the weekday demand pattern of 2019 during thee same time period (Supplementary Figure S6). Immediately after the lockdown (around week 12, as most global lockdown measures happend in this week), the energy demand profiles started shifting to more weekend-like profiles, with lower and delayed morning peaks (Supplementary Figure S6). These shifts indicate that the "working-from-home" thus remote working style had driven the change in energy demand pattern.

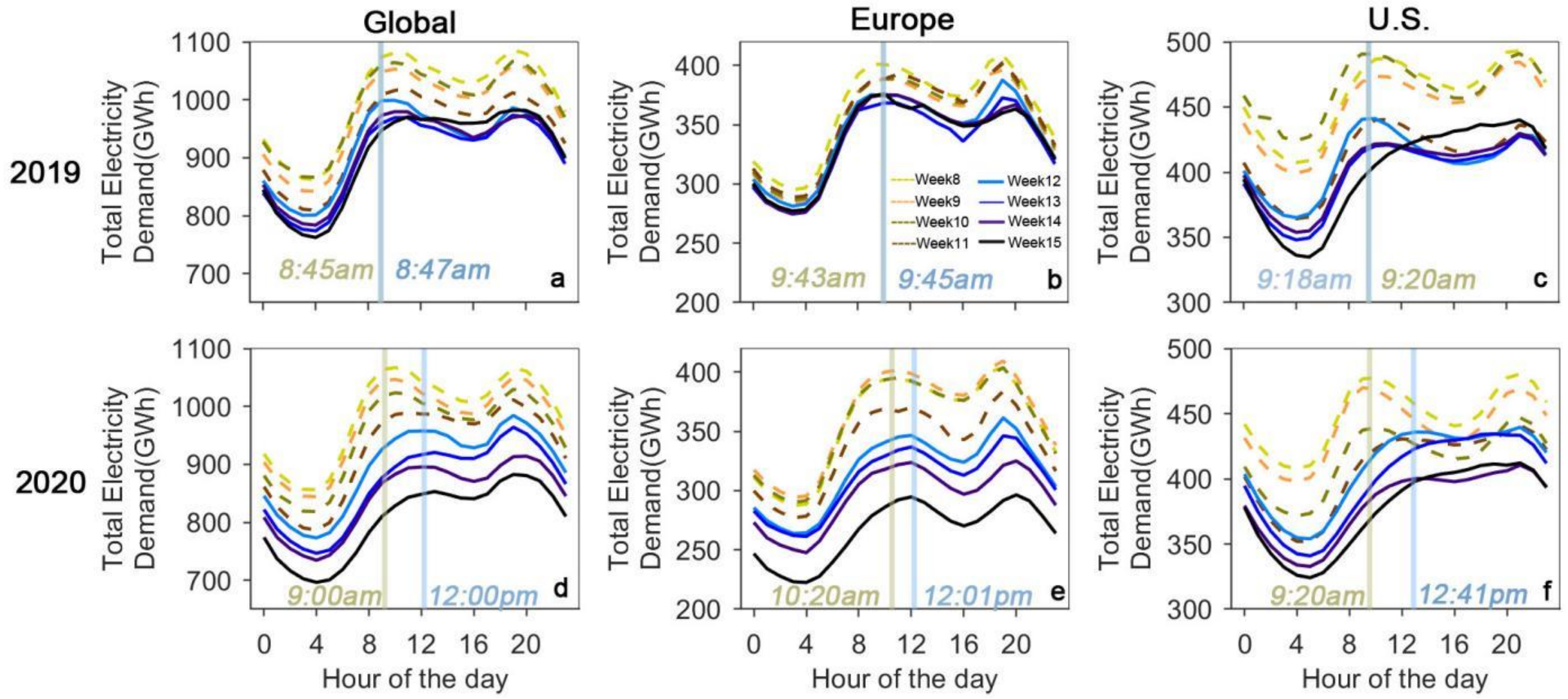

**Figure 3 | Diurnal Pattern of Total Energy Demand at Global level, of Europe and the U.S.** Each line represents a weekly average of the diurnal pattern of the total energy demand. Global pattern includes data from EU27&UK, U.S., Japan and Brazil. Dashed line indicates week8-week11, which are the four weeks prior to lockdown measures taking place in 2020. Lockdown measures in most countries/regions took place in week 12, 2020. Solid lines indicate the four weeks (week12 - week 15) when diurnal profiles were under the impact of the COVID-19 pandemic in 2020. For 2019, solid lines and dashed lines indicate the same time period (but without the influence of COVID-19). Vertical lines indicate the times for morning peaks in week8-week11 (prior-COVID for 2020, green line) or in week12 - week 15 (post-COVID for 2020, blue lines). For diurnal patterns of other countries, see Supplementary Fig. S5.

## New electricity demand pattern as a driving force for increased share of low carbon sources in the electricity system

A shift towards higher energy consumption around mid-day, while lower towards morning and night time, likely have facilitated the increase of low-carbon energy (particularly solar) share in the energy mix. Traditionally, power system is known to have to ramp up the thermal electricity to cope with the morning peak (Fig. 3 and thermal electricity profiles in Supplementary Fig. S7). Because of the lowered and delayed morning peak, and the generally lower daytime consumption, thermal power has lowered its morning peak as well. This is reflected by the significant decline in the morning ramp rate change of the "duck curve" (Fig. 4), or the electricity demand less solar electricity generation[16]. Its characteristic morning ramp and evening ramp (climb from midday low point to evening peak) have both changed significantly following the lockdown measures. Compared to pre-COVID days, and to the same time period of reference year 2019, the post-COVID morning and evening ramp rates have declined at global scale (Fig. 5). The evening ramp rate decline is much more obvious for the U.S., caused by the increased midday consumption.

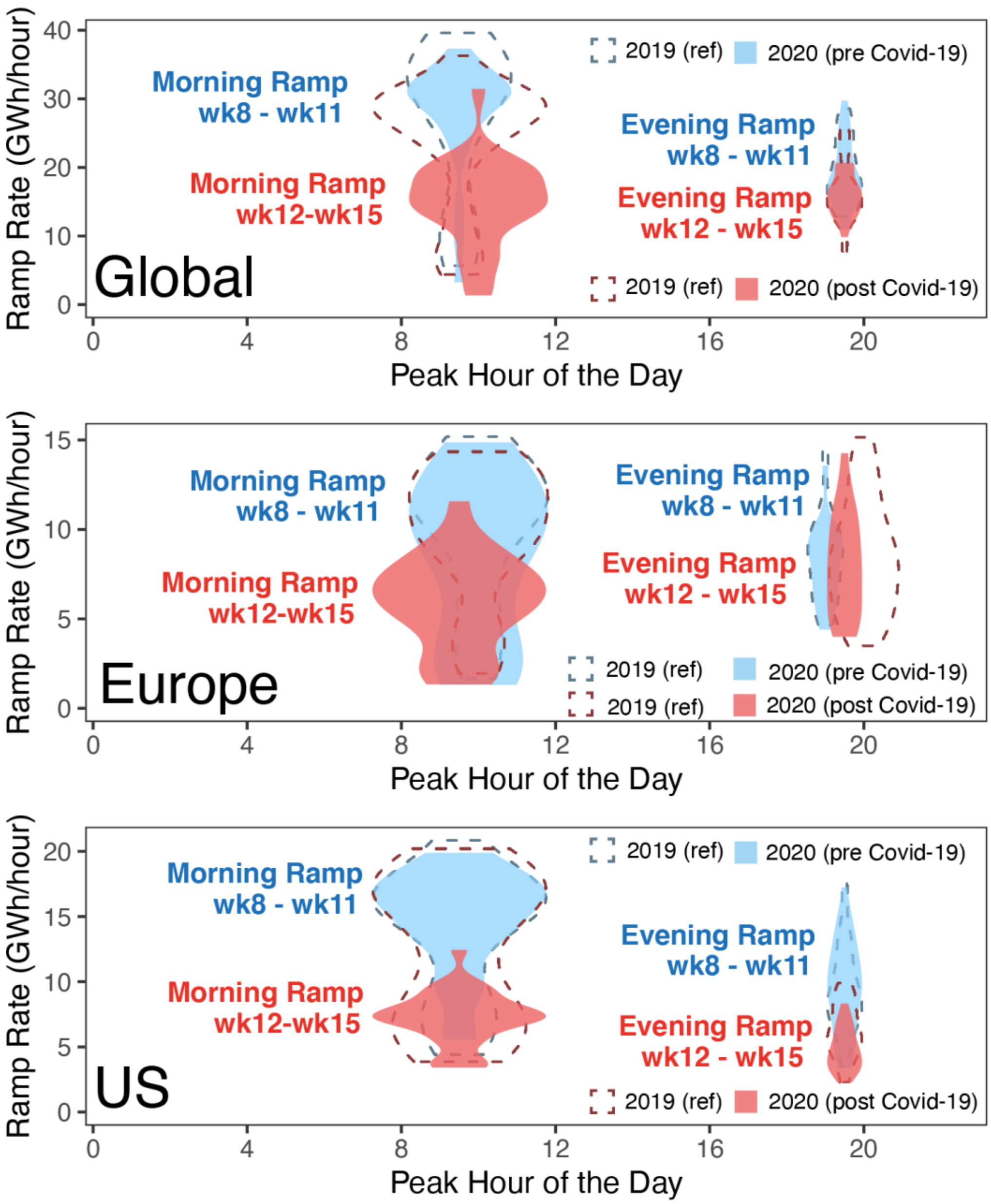

**Figure 4 | Non-solar electricity consumption ramp rate change in 2019 and in 2020 for the periods before and after lockdowns** Summary of changes in "duck curves" (i.e. solar-excluded electricity consumption, see Supplementary Fig. S6). Morning ramp rate is computed by dividing the morning ramp height by the hours this ramp lasted (time taken for the energy consumption to reach from the lowest point to the morning peak). Morning ramp height is computed as the energy consumption difference between morning peak (highest energy consumption before midday) and the midnight lowest point (usually around 4AM). Evening ramp rate is

computed by dividing the evening ramp height by the hours this ramp lasted. Evening ramp is defined as the difference in energy consumption between midday lowest demand and the evening peak (highest consumption during after midday, usually occurs around 7-8PM).

The diurnal patterns of each electricity sourses were affected differently during the COVID-19 pandemic period (Fig. 6). The most noticeable change is that thermal power from coal and gas declined more during day time, as a result of adjusting to the lowered morning consumption peak demand, and of adjusting to a higher day-time renewable energy production. In the U.S. in 2020, we observe increases in gas power compensating for decreased coal power in the U.S. during the afternoon in March, and during the night in April and May (Fig. 5). In Europe, there was a similar increase of gas power during morning hours in March when the weather was colder than in 2019, and a strong morning decrease in April and May that were warmer than in 2019 (Supplementary Table S3). As solar power peaking around midday was much above average in 2020, coal energy also declined during midday (Fig 5). Similar changes are also detected at smaller spatial scales (national and sub-national, Supplementary Fig. 4). Although coal generation was already lower than in 2019 for Europe and the U.S. in January and February, that pre-COVID deficit was uniformly distributed throughout the day (Fig. 6). After the lockdowns, this pattern changed to more decline during the day, especially around 8 AM (coal power in the U.S. and in coal and gas power combined in Europe). This could be explained by the flattened morning peaks in energy demand profiles (Fig. 3). As there is less electricity demand at around 8AM, likely caused by life styles changing towards widely practiced remote working.

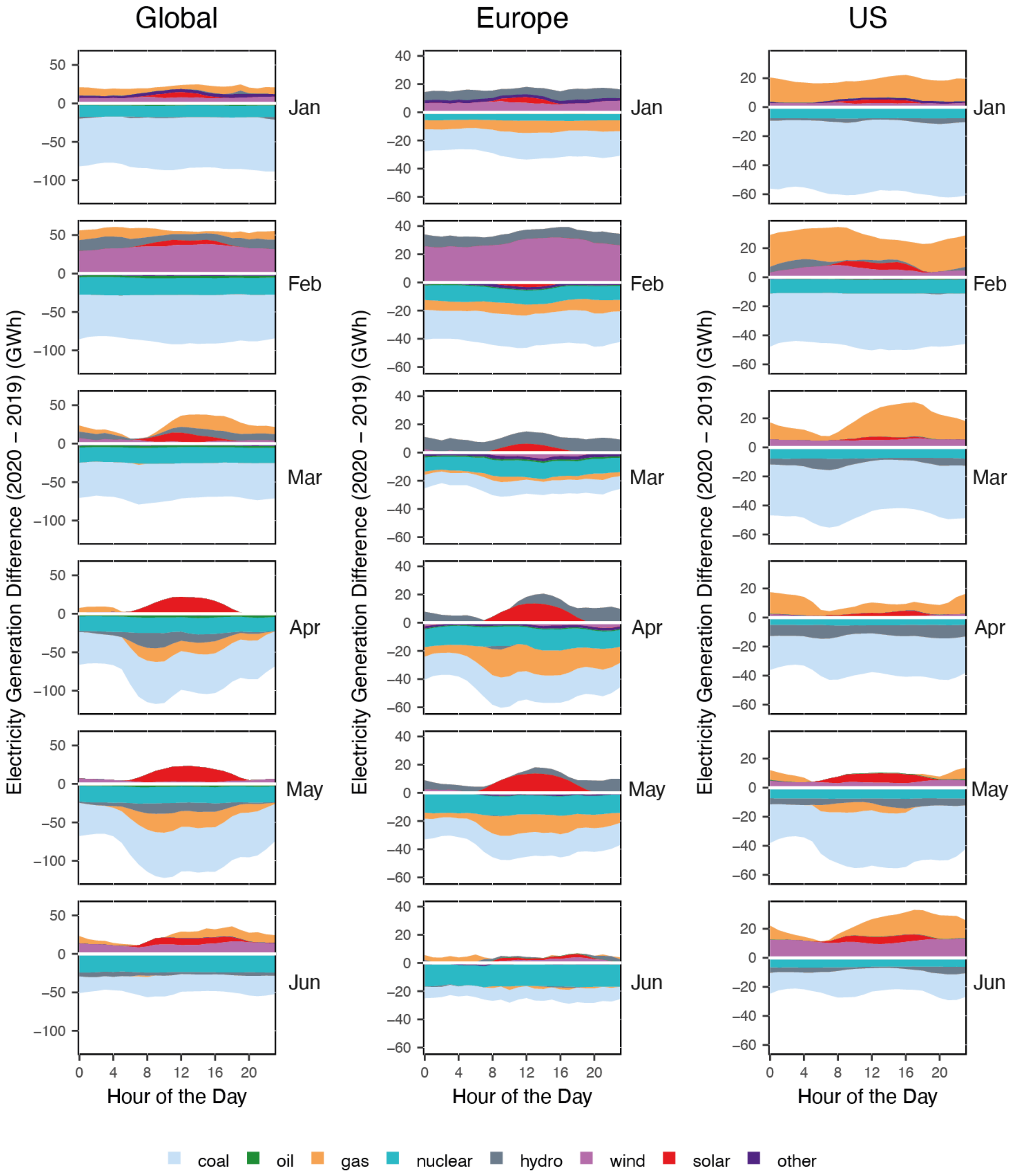

**Figure 5 | Source-specific Power generation Difference between 2020 and 2019, shown as monthly averaged diurnal pattern, for month January to June.** Positive values indicate an increase of power generation at a given time of the day in year 2020 compared to 2019, white negative values indicate a decrease. The profiles are computed by taking the difference between the monthly averaged diurnal patterns of energy generated by each source from year 2020 and year 2019. Global profiles include data from Europe (EU27&UK), US, Japan and Brazil. Regional comparisons of other major countries/regions see Supplementary Fig S4.

## Rebound of fossil fuel power generation

The transient drop of carbon intensity in the power sector during the first half of 2020 appears to result from a combination of three factors: a lower demand from declined industrial and commercial activities; a market-driven substitution from coal to gas especially in the U.S.; and an exceptional supply of both solar and wind power occurring at the same time in Europe, the US and China, although during different months in each of these countries/regions. However, fossil fuel power generation and emission intensities have rebounded in many countries/regions. Since May 2020 this rebound was mainly contributed by China and countries with released lockdown measures. The carbon intensity in May reached back to its level of 2019 at the same time of year. To gain insights into the drop of GHG emissions from the power sector (Fig1), we decomposed monthly emission changes from 2016 to mid-2020 into the contributions of changes in total power generation (activity) versus changes in energy sources mix (structure). Results (shown in Fig6, given as decomposed contributions to change in GHG emissions, with reference to the same time period in 2016) suggest that from January 2017 to January 2019, increased activity (increased power generation) outpaced the improvement in structure (increased low carbon sources' share) globally. In addition, the global decomposition is similar to that of China, the largest emitter. From January, 2019 to the beginning of the first lockdown in 2020 (in Wuhan, China), improved structure is noticeable, contributed mainly by Europe and secondarily by the U.S.. After the first lockdown in China, the sharp decline of global emissions is explained by both decreased activity (59%) and improved share of low carbon sources in energy mix (41%). There was no clear improvement in energy mix related to the 2020 lockdown in China compared to 2019, a higher low carbon share in energy mix in Europe than in 2019, while in the U.S., the decreased emissions is mainly explained by decreased activity (Fig. 6).

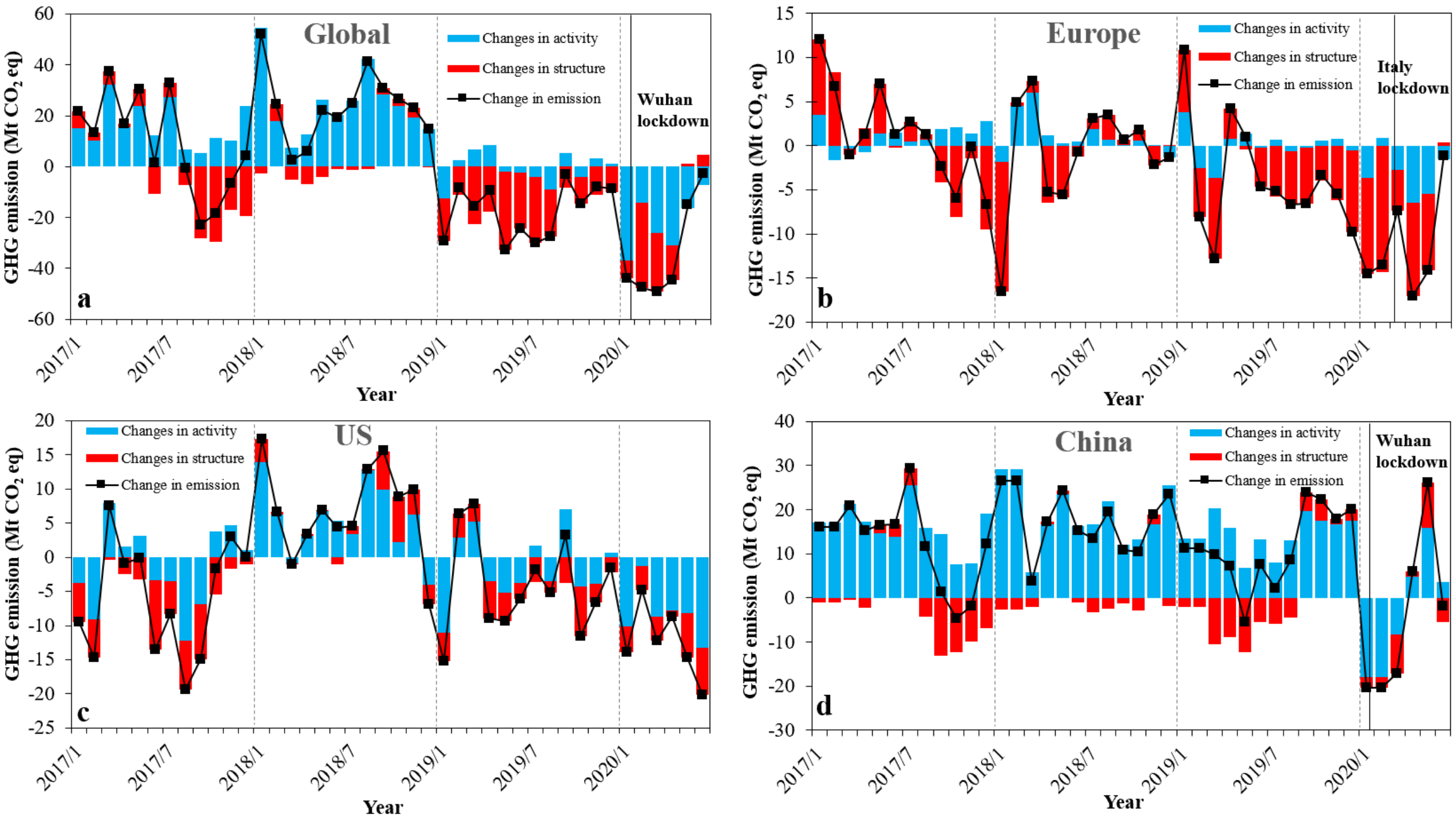

**Figure 6 | Decomposed changes in regional emissions.** Logarithmic Mean Divisia Index (LMDI) decomposition analysis of Greenhouse Gas (GHG) emissions from global power generation. Monthly GHG emission data is compared with the same time period of the previous year to remove the inherent seasonal variations. Change in activity refers to the change in gross electricity generation. Change in structure refer to the change in each energy source sector's share in energy mix. Change in emission refers to the difference between the GHG emissions from the final month and the GHG emissions from the same month in previous year.

# Discussion

During the COVID-19 pandemic, the carbon intensity of the global power sector reached historic lows. Although there were many factors underlying this decarbonization, three drivers are most prominent: First, electricity generation from solar and wind--which have been growing strongly for years--reached new highs in many regions. Meanwhile, sudden and substantial drops in electricity demand were disproportionately met by turning down fossil fuel-fired power plants. The decrease in fossil fuel use makes economic sense despite low oil and gas prices given the near-zero marginal costs of solar and wind electricity, but the greater use of variable renewable energy sources--and particularly solar--was further bolstered by shifts in the diurnal pattern of energy demand, with lower ramp rates and peak loads that were lower and nearer to the middle of the day (i.e. loads that were easier for the available renewable sources to meet).

We analyzed the impact of COVID-19 on global power system with a novel near-real-time country specific, source specific data-set. With this approach, we are able to detect the immediate and long-term impact of lockdown measures on the global power generation and consumption system. Our results show that as individual behaviours change in response to lockdown measures, the diurnal cycle of power systems distorted strongly in 2020, which favored solar power by moving the peak demand closer to the peak of daytime radiation. However, there are still uncertainties remained in our results. Although we have provided a fairly comprehensive dataset for the purpose of this study, there are still limitations in daily availability in general. High-time resolution and high quality sub-hourly data remains a challenge to acquire for some parts of the world. Secondly, for the GHG emission estimation we applied an IPCC life-cycle emission factor based approach. Uncertainties may rise from the emission factors, which are assumed to be universal for a given power source sector. In future studies, country-specific and time-specific emission factors may be applied to reduce these uncertainties.

However, in China and other countries, the end of lockdowns was followed by a rapid rebound of fossil power generation. Although our analyses demonstrate the potential of existing renewable electricity sources to be robust and reliable sources of power, the return to normal levels and timing of electricity demand are dragging the carbon intensity of electricity systems around the world back up. Yet the transient changes during the pandemic suggest that focused investments under a green recovery could be influential. In particular, the economic viability of coal-fired power seems near a tipping point in many regions[17]. such that modest additional deployment of renewables and energy storage technologies (the latter of which could act to shave and redistribute intraday peaks) might trigger large and permanent decreases in the shares of coal power and thus the carbon intensity of electricity.

# Methods

**Data availability and process.** We focused on six countries/regions, i.e., China, India, United States, Europe (EU27 & UK), Japan and Brazil in this research. The total electricity generation of these regions accounts for ~68% of the global energy production (see Supplementary Table S1). We compiled and estimated hourly and sub hourly power generation and demand data with publicly available datasets (Table 1). Time zone-corrected electricity data, which are collected at the sub-regional or sub-national levels such as in the U.S. and Europe, are then combined into the national or regional total.

We defined eight major energy source categories  in this study: coal, gas, oil, nuclear, hydro-power, wind, solar and other (note that "other" represents the combined category for biomass, geothermal and other renewables). For data sources in which sub-categorical data are provided, we aggregated these sub-categories to one of the eight categories summarized above. For example, onshore wind and offshore wind are both included in wind energy in this study. In addition, thermal and low carbon energy sources are defined based on the energy sources' carbon intensity. Thermal energy includes energy produced by coal, gas and oil. Low carbon energy includes energy produced by nuclear, hydro-power, wind, solar, biomass, geothermal and other renewables. According to data availability (Table 1), sub-categorical data were re-grouped to the eight major energy source categories described above. The detailed contents of each category are described as following:

- Coal includes: Coal (EIA, CEC, POSOCO, BP), Lignite (POSOCO), Fossil Brown coal/Lignite (ENTSO-E), Fossil Hard coal (ENTSO-E);
- Oil includes: Oil (IEA, CEC, BP), Petroleum (EIA), Fossil Oil (EIA), Fossil Oil shale (ENTSO-E); Naptha & Diesel (POSOCO)
- Gas includes: Natural Gas (EIA, CEC, BP), Fossil Coal-derived gas (ENTSO-E), Fossil Gas (ENTSO-E);

- Nuclear includes: Nuclear (EIA, ENTSO-E, IEA, OCCTO, ONS, CEC, POSOCO, BP);
- Hydro-power includes: Hydro (EIA, OCCTO, ONS, CEC, POSOCO, BP), Hydro Pumped Storage (ENTSO-E), Hydro Run-of-river and poundage (ENTSO-E), Hydro Water Reservoir (ENTSO-E), Pumped-storage hydroelectricity (OCCTO);
- Wind includes: Wind Offshore (ENTSO-E), Wind Onshore (ENTSO-E), Wind (EIA, NBSC, CEC, OCCTO), Wind (regulated, OCCTO);
- Solar includes: Solar (EIA, ENTSO-E,ONS, CEC, POSOCO, BP), Photovoltaic (OCCTO), Photovoltaic (regulated, OCCTO);
- Biomass, Geothermal, and Other Renewables: Biomass (IEA, OCCTO, BP), Combustible Renewables (IEA), Geothermal (ENTSO-E, OCCTO, IEA, BP), Other renewables (ENTSOE, IEA), Waste (ENTSO-E)

For countries/regions where high-temporal resolution data from main source categories are not available, or where several categories are combined, energy source category specific data are estimated or disaggregated using the ratio from available monthly and/or yearly data. The Code for processing the data will be made available on Github.

**Table 1. Data availability** | The data availability for the studied countries/regions which were presented in Results are summarized in this table. In total, 30 countries are included. Data Platform provides the official names of platforms where data for this study are acquired. URL provides the link to access the data files for this study. "Published since" provides information on the earliest available data from each platform. Spatial resolution refers to the aggregated and sub areas at which data are available. Temporal Resolution provides information for the resolution of data which can be directly downloaded (without further processing) from the platform. Production Types list all the energy generation source types (major categories and subcategories) that are included in this study.

| Countries/Regions | Data Platform | URL | Published since | Spatial Resolution | Temporal Resolution | Production Types |
|---|---|---|---|---|---|---|
| U.S. | EIA Hourly Electric Grid Monitor | https://www.eia.gov/beta/electricity/gridmonitor | 2018/7/1 | 13 sub-regions of the lower 48 states: Southeast, Midwest, Northwest, Southwest, California, Carolinas, Texas, Florida, New England, New York, Mid-Atlantic, Central, Tennessee. | Hourly, Daily | Coal, Petroleum, Natural gas, Wind, Solar, Hydro, Nuclear, Other |
| India | Power System Operation Corporation Limited (POSOCO) | https://posoco.in/reports/daily-reports/ | 2013/4/1 | National total , and 5 sub regions: NR, WR, SR, ER, NER | Daily | Coal, Lignite, Hydro, Nuclear, "Gas, Naptha & Diesel", "RES (Wind, Solar, Biomass & Others)" |
| Brazil | Operador Nacional do Sistema Elétrico (ONS) | http://www.ons.org.br/Paginas/resultados-da-operacao/historico-da-operacao/geracao_energia.aspx | 1999/1/1 | National total, and all 26 states of 4 regions, i.e., Nordeste (Northeast), Norte (North), Sudeste/Centro-Oeste (southeast/midwest), and Sul (south). | Hourly, Daily, Monthly, Annually | Hydro, Wind, Nuclear, Solar, Thermal |
| Europe (EU27&UK) | ENTSO-E Transparency Platform | https://transparency.entsoe.eu/generation/r2/actualGenerationPerProductionType/show | 2014/1/1 | 35 countries: Armenia, Austria, Azerbaijan, Belgium, Bosnia and Herz Bulgaria, Croatia, Cyprus, Czech Republic, Denmark, Estonia, Finland, France, Georgia, Germany, Greece, Hungary, Ireland, Italy, Latvia, Lithuania, Montenegro, Netherlands, North Macedonia, Norway, Poland, Portugal, Romania, Serbia, Slovakia, Slovenia, Spain, Sweden, Switzerland, Turkey, United Kingdom | 15 min interval ~ 1 hour interval | Biomass, Fossil Brown coal/Lignite, Fossil Coal-derived gas, Fossil Gas, Fossil Hard coal, Fossil Oil, Fossil Oil shale, Fossil Peat, Geothermal, Hydro Pumped Storage, Hydro Run-of-river and poundage, Hydro Water Reservoir, Marine, Nuclear, Other, Other renewable, Solar, Waste, Wind Offshore, Wind Onshore |
| Japan | Organization for Cross-regional Coordination of Transmission Operators (OCCTO) | https://www.occto.or.jp | 2018/4/1 | National total from 10 electricity providers: Hokkaido Electric Power Network, Tohoku Electric Power Network, Tokyo Electric Power Company, Chubu Electric Power Grid, Hokuriku Electric Power Company, Kansai Electric Power, Chugoku Electric Power Transmission & Distribution, Shikoku Electric Power Company, Kyushu Electric Power Transmission & Distribution, and Okinawa Electric Power Company | Hourly | Nuclear, Thermal, Hydro, Geothermal, Biomass, Photovoltaic, Photovoltaic (regulated), Wind, Wind (regulated), Pumped-storage hydroelectricity |
| China | National Bureau of Statistics, China (NBSC) | https://data.stats.gov.cn/ | 1989/2/1 | National total, and 31 Provinces in mainland China | Monthly, Annually | Thermal, Hydro, Nuclear, Wind, Power |
| China | China Electricity Council (CEC) | https://english.cec.org.cn/menu/index.html?251 | 2013/11/1 | National total | Monthly, Quartly, Annually | Thermal, Hydro, Nuclear, Wind, Solar, Coal (estimated from Coal Power Plant Installed Capacity and Coal Power Plant Operation Time), Natural Gas (estimated from Natural Gas Power Plant Installed Capacity and Natural Gas Power Plant Operation Time) |
| World | International Energy Agency (IEA) | https://www.iea.org/reports/monthly-electricity-statistics | 2000/1(monthly); 1971(yearly) | National total, for OECD Member Countries and a selection of IEA Association Countries and other economies | Monthly | Conventional Thermal (Coal, Oil, Natural Gas, Combustible Renewables, Other Combustibles), Nuclear, Hydro, Wind, Solar, Geothermal, Other Renewables |
| World | Statistical Review of World Energy by BP | https://www.bp.com/en/global/corporate/energy-economics/statistical-review-of-world-energy.html | 1965 (Low carbon sources); 1985(Fossil sources) | World total, Continental total and National total, for | Yearly | Coal, Oil, Gas, Nuclear, Hydro, Wind, Solar, Geothermal, Biomass and Other |

**Least-Squares Regression Method**

We estimated the temperature induced power generation change by establishing a linear regression model between the daily power generation in year 2018 & 2019 with the temperature on the corresponding day, for each country, the following model (Eq.1) is established:

$$E_c = \alpha_c T_c + \beta_c \quad \text{(Eq.1)}$$

For each country c, the  and  are estimated by minimizing the Root Mean Squared Error (RMSE ), expressed as Eq.2:

$$\text{RMSE} = \left(\sum_{i=1}^{n} \frac{1}{n} (\widehat{E_{c,i}} - E_{c,i})^2\right)^{\frac{1}{2}} \quad \text{(Eq.2)}$$

Where $\widehat{E_{c,i}}$ is the calculated power generation for country c, day i with Eq.1, and the $E_{c,i}$ is the actual power generation for country c, day i. The criteria for establishing a meaningful linear regression model between the power generation and temperature is that the linear relationship between power generation (Ec) and temperature (Tc) is strong enough, i.e. temperature variation can explain the majority of the change in electricity generation. Therefore, we excluded public holidays, when social activities contribute substantially to power generation change. The criteria for a medium to strong linear relationship is that the R-squared value of the established model is above 0.6. Once the regression model is established, we calculated the theoretical power generation change with the temperature difference between 2020 and

2019 on the same day.

**Emission Calculation.**

The emissions from the energy mix are calculated by multiplying the activity data (i.e., electricity production) and the emission factors by energy types (Eq. 3)[18].

$$\text{GHG Emissions} = \Sigma(\text{Activity Data}_i \times \text{Emission Factor}_i) \qquad \text{(Eq. 3)}$$

In this study, we analyze the GHG emissions from the whole life cycle of electricity production. The life cycle GHG emissions are associated with the upstream energy consumption and the embodied energy use for building power stations and operation[19], which are frequently used to analyze the environmental cost and carbon footprint of different power generation technologies[20]. Here, We use the life cycle emission factors by different technologies from the IPCC (2015)[21] (see Supplementary Table S2) to assess the GHG emissions over the whole life cycle per unit electricity generated. Different technologies show large discrepancies on their life cycle emissions. For example, nuclear energy is one of the lowest GHG emission technologies. The life cycle GHG emissions of nuclear generated electricity is only 1/30 of the emissions caused by coal generated electricity.

The carbon intensity of energy mix is calculated by the following equation (Eq. 4), defined as the GHG emitted per unit of electricity produced:

Carbon Intensity= GHG emissions/Electricity Production (Eq. 4)

**Index Decomposition Analysis.**

The Logarithmic Mean Divisa Index (LMDI) decomposition method was proposed to solve different decomposition problems in 1998[21]. Here, LMDI is adopted to perform the residual-free decomposition of the factors affecting GHG emissions. The expression is shown in Eq. 5) (adopted from literature[22,23]):

$$E = \sum_i E_i = \sum_i Q \frac{Q_i}{Q} \frac{E_i}{Q_i} = \sum_i Q S_i I_i \qquad \text{(Eq. 5)}$$

Where *E* is the total GHG emission (expressed as CO2-eq), Q is the gross generation from energy, and *Si* and *Ii* are the shares of electricity *i* generation and emission factor (CO2-eq/generation) of electricity source *i*, respectively. Taking 2016 as the baseline year (each month in 2016 as the baseline months), thus the contribution of each driving factor causing the emission change (difference in GHG emissions between the t$^{th}$ month and the base month) can be expressed as the following equation (Eq. 6):

$$\Delta E_t = E^t - E^0 = \Delta E_{act} + \Delta E_{str} + \Delta E_{int} \quad \text{(Eq. 6)}$$

Where ΔE is the difference between the GHG emissions from the t[th] month ($E^t$) and the GHG emissions from the baseline month ($E^0$). In this study, by developing the time-series manner, $E^0$ from January 2016 to June 2019 and $E^t$ from January 2017 to June 2020 are considered with one month step. $\Delta E_{act}$, $\Delta E_{str}$, and $\Delta E_{int}$ represent the activity effect, the energy structure effect, and the emission intensity effect, respectively. Activity effect is indicated by gross power generation, energy structure effect is indicated by energy's share of power generation, and emission intensity effect is indicated by energy's emission factor, which is shown in Eq.7 - Eq.10:

$$\Delta E_{act} = \sum_i w_i \ln(\frac{Q_i^t}{Q_i^0}) \quad \text{(Eq.7)}$$

$$\Delta E_{str} = \sum_i w_i \ln(\frac{S_i^t}{S_i^0}) \quad \text{(Eq.8)}$$

$$\Delta E_{str} = \sum_i w_i \ln(\frac{I_i^t}{I_i^0}) \quad \text{(Eq.9)}$$

$$w_i = \frac{E_i^t - E_i^0}{\ln E_i^t - \ln E_i^0} \quad \text{(Eq. 10)}$$

Where $Q^t$, $S^t$, and $I^t$ are the power generation, energy structure, and emission factor of the t[th] month, respectively. $Q^0$, $S^0$, and $I^0$ are the power generation, energy structure, and emission factor of the baseline month, respectively.

**Data Availability Statement**
All data generated or analyzed during this study are included in this article, source data are available from the corresponding author upon reasonable request.

**Code Availability**
All code generated during and/or analyzed during the current study are available from the corresponding author upon reasonable request

**Competing Interests statement**
Authors declare no competing interests.

**Author contribution**
Author contributions: Z.L. designed the paper and prepared the initial draft; B. Z. coordinated the data preparation with contributions from C. L., H.Z., P.K., Y. C., Z.D., D. C., T. S., X. D., J. T.,B. Z.; S. D. designed the Fig 4. Z. L., B. Z., S. D., P. C. contribute to paper writing with help for all other authors.

**Acknowledgement**

ZL acknowledges funding from the National Natural Science Foundation of China (grant 71874097 and 41921005), Beijing Natural Science Foundation (JQ19032) and Qiushi Foundation.